\documentclass[a4paper,11pt]{article}
\usepackage{jcappub}
\usepackage{aas_macro}

\usepackage[T1]{fontenc}
\usepackage{graphicx}
\usepackage{orcidlink} 
\usepackage{bm}
\usepackage{epsfig}
\usepackage{amsfonts}
\usepackage{amssymb,amscd}
\usepackage{subfigure}
\usepackage{xcolor}
\usepackage{comment}
\usepackage{amsmath}
\usepackage{graphicx}
\usepackage{dcolumn}
\usepackage{hyperref}

\begin{document}

\title{
    Probing the sensitivity of CTAO-N LSTs observations at large zenith angles
    to the multi-TeV gamma-ray emission
    from the inner 10 parsecs of the Galactic Center
}

\author[a]{Shotaro Abe \orcidlink{0000-0001-7250-3596} }
\author[a,b,c]{Tomohiro Inada \orcidlink{0000-0002-6923-9314}}
\author[d]{Emmanuel Moulin \orcidlink{0000-0003-4007-0145}}

\affiliation[a]{Institute for Cosmic Ray Research, University of Tokyo, 5-1-5, Kashiwa-no-ha, Kashiwa, Chiba 277-8582, Japan}
\affiliation[b]{Kyushu University, Nishi-ku, 819-0395 Fukuoka, Japan}
\affiliation[c]{CERN, CH-1211 Geneva 23, Switzerland}
\affiliation[d]{Irfu, CEA Saclay, Université Paris-Saclay, F-91191 Gif-sur-Yvette, France}

\emailAdd{shotaro@icrr.u-tokyo.ac.jp}
\emailAdd{tomohiro.inada@cern.ch}
\emailAdd{emmanuel.moulin@cea.fr}

\date{\today}

\abstract{
    Observations of the Galactic Center using Imaging Atmospheric Cherenkov Telescopes (IACTs), such as H.E.S.S., MAGIC, and VERITAS, have revealed a very-high-energy (VHE, $\gtrsim 100$ GeV) gamma-ray source, HESS J1745$-$290, aligned with the dynamical center of the Milky Way.
    This source shows point-like emission ($\lesssim 0.1^\circ$) and a strong suppression in its energy-differential spectrum in the ten TeV energy regime, modeled well by a power-law with an exponential cutoff.
    The origin of this emission is debated, with candidate emission scenarios including dark matter annihilations, millisecond pulsars in the central stellar clusters, and hadronic interactions in the vicinity of Sagittarius A*. 
    Deriving the sensitivity to these spectral models is key to discriminating the physical processes at work.
    We show that combining H.E.S.S., MAGIC, and VERITAS archival data can  well described the observed emission by a power-law with an exponential energy cutoff within the present uncertainties.
    Given the near advent of the array of the Large-Sized Telescopes (LSTs) at CTAO-N,
    we timely simulate realistic upcoming observations of the central emission by
    the CTAO-N four-LST array, to derive the sensitivity to resolve the sharpness of the spectral energy cutoff.
    We find that 500 hours of four-LST observations taken at large zenith angles, possibly accumulated over several years, can significantly discriminate the dark-matter emission scenario from the leptonic and hadronic ones. 
    Also, a preliminary 3$\sigma$ hint for such discrimination could emerge within the first year. 
    We demonstrate, for the first time, that CTAO-N is able to provide new insights on differentiating among the above-mentioned emission senarios in the next several years. 
}

\maketitle

\section{Introduction}
\label{sec:intro}

    A very-high-energy (VHE, $E \gtrsim$ 100 GeV) gamma-ray emission spatially coincident with the supermassive black hole Sagittarius A* (Sgr A*) in the Galactic Center (GC) region has been detected by Imaging Atmospheric Cherenkov Telescopes (IACT) such as Whipple~\cite{Kosack:2004ri}, H.E.S.S.~\cite{Aharonian:2004wa}, MAGIC~\cite{Albert:2005kh} and VERITAS~\cite{Archer:2014jka}. 
    Further observational campaigns revealed the emission to be compatible with a point-like source 
    ($\lesssim$0.1$^\circ$) with an energy-differential spectrum following a power-law exhibiting a strong suppression around 10 TeV, for which a power-law parametrization with an exponential energy cut-off provides a significantly preferred fit to the data compared to the power-law one~\cite{Aharonian:2009zk,Abramowski:2016mir}.

    Spatial studies have been conducted by H.E.S.S. to pinpoint the underlying astrophysical counterparts. 
    Despite the limited angular resolution of IACTs and present photon statistics, careful astrometric pointing corrections were developed and enable to exclude the supernova remnant Sgr A East as a dominant contribution of the observed emission, leaving Sgr~A* and the pulsar wind nebula candidate G359.95-0.04 as potential counterpart~\cite{2010MNRAS.402.1877A}.
    Variability searches have been carried out in the VHE flux since Sgr A* is the source of bright and frequent X-ray and infrared flares~\cite{2001Natur.413...45B}, and quasi-periodic oscillations (QPO) on time scales of 100-2250 s may have been detected~\cite{2004A&A...417...71A}. 
    However, no variability or QPOs have been found in flux lightcurve by VERITAS, MAGIC and H.E.S.S.~\cite{Aharonian:2009zk} nor flaring activity during simultaneous H.E.S.S. and Chandra observations of Sgr~A*~\cite{Aharonian:2008yb}.

    Several scenarios have been proposed to explain the observed VHE emission including interaction of cosmic-ray protons accelerated in the vicinity of Sgr A* interacting in the ambient medium producing gamma rays via $\pi_0$ decay~\cite{Aharonian:2004jr,Liu_2006,Yan-Ping_Wang_2009}, inverse Compton scattering of electrons accelerated at the wind termination shock within a few Schwarzschild radii~\cite{Atoyan:2004ix} or in the pulsar wind nebula G359.95-0.04~\cite{Hinton_2007}, off the radiation field, a spike of annihilating dark matter (DM)~\cite{PhysRevD.86.083516,Cembranos:2012nj}, and millisecond pulsars (MSP) in the central stellar cluster Sgr A~\cite{Bednarek:2013oha}. 
    However, since its detection in 2004, its origin is still a mystery: any potential association to an astrophysical counterpart cannot be unambiguously claimed.

    Spectral gamma-ray template for HESS J1745-290 can be expressed by a super-exponential energy-cutoff power law (SEPL) given by 
    \begin{align}
            \Phi(E) = \Phi_0 \times \left(\frac{E}{1\,\mathrm{TeV}}\right)^{-\Gamma} \times \exp \left( -\frac{E}{E_c} \right)^{\beta}
            \label{eq:general_ECPL} \, ,
    \end{align}
    where $\beta$ is the shape parameter in the energy cutoff region and $E_c$ is the cutoff energy. $\Gamma$ and $\Phi_0$ stand for the spectral index and the flux normalization, respectively.
    A DM-induced model can be well described by SEPL parametrization with $\beta > 1$, and a MSP-induced model by SEPL with $\beta = 1$ as will be shown in the following. A proton-induced model can be characterized by a power-law with an exponential cutoff with $\beta < 1$~\cite{PhysRevLett.111.071302}.
    The shape of the energy cutoff therefore carries crucial information on the underlying gamma-ray emission mechanism. 
    As the sharpness of the suppression at several TeV energies in the differential energy spectrum of HESS J1745-290 may be a powerful signature of the underlying VHE gamma-ray production processes, we will explore the sensitivity of the forthcoming observations with the Cherenkov Telescope Array Observatory (CTAO) to probe emission scenarios of the VHE gamma-ray emission at the Galactic Center (GC). 
    With the imminent advent of four Large-Sized Telescopes in the Northern site of the CTA Observatory (CTAO) within the next few years~\cite{CTALSTProject:2023vhk}, we timely forecast how much the upcoming observations can provide new insights into the understanding of the nature of VHE GC source. 
    We investigate how the increased photon statistics and control of the systematic uncertainty to the level achieved by current measurements, can help to discriminate among the discussed emission scenarios via spectral measurements.

    The Cherenkov Telescope Array Observatory (CTAO) is the next-generation IACT observatory, situated at two sites to extensively cover the sky: the Northern site, CTAO-N, in La Palma, Canary Islands, and the Southern site, CTAO-S, in Paranal, Chile. CTAO is composed of three types of telescopes: Large-Sized Telescope (LSTs, 23 m diameter), Medium-Sided Telescope (MSTs, 12 m diameter), and Small-Sized Telescopes (SSTs, 4.3 m diameter). 
    The \textit{Alpha} configuration of CTAO-S is expected to have 14 MSTs and 37 SSTs , and that of CTAO-N is 4 LSTs and 9 MSTs~\cite{cta_observatory}. 
    The first LST in CTAO-N has already attained first light and been observing the Galactic Center region \cite{CTALSTProject:2023wzu}. 
    The remaining other three LSTs in CTAO-North are under construction, and the operation of 4LST array is planned to start operating in 2026 \cite{CTALSTProject:2023vhk}. 
    %
    Since IACTs uses the atmosphere as a calorimeter, the zenith angle (or zenith distance, Zd) at which telescopes point, significantly influences observational performance, as the distance to the shower axis varies accordingly.
    Observations at low zenith angles are generally considered the standard operational mode for IACTs, due to the reduced thickness of the atmospheric layer through which the air showers propagate.
    Owing to its geographical location, CTAO-S is anticipated to offer an unprecedentedly enhanced view of the Galactic Center (GC) region, sometimes regarded as the only viable observational site towards this target.
    In contrast, from the CTAO-N site, where the GC culminates at a zenith angle of approximately 58$^{\circ}$, the so-called large-zenith-angle (LZA) observation technique must be employed.
    Importantly, while this method typically entails a compromise in terms of energy threshold and resolution, it substantially increases the effective collection area for VHE gamma rays up to an order of magnitude.
    As a result, LZA observations enable an alternative data collection even with instruments primarily optimized for lower energies, including the LSTs, achieving a collection area of $\sim$ 9.0 $\times 10^5$\ m$^2$ at 10 TeV.
    Given that exposure time becomes increasingly crucial in the VHE regime, the early accumulation of data and a strategic observational approach are essential.
    In light of this context, and in addition to the future CTAO-S array, it is timely to evaluate the spectral sensitivity of the four-LST sub-array to distinguish among potential physical models devised to explain the central VHE emission, given its completion being planned for 2026.

    The paper is organized as follows. 
    Section~\ref{sec:models} summarizes spectral models designed to explain the VHE gamma-ray emission at the GC region in light of the latest measurements performed by currently operating IACTs: 
    a spike of annihilating DM around the supermassive black hole Sgr~A*; relativistic protons accelerated in the vicinity of Sgr~A* interacting in the ambient material; and a population of millisecond pulsars in the central stellar cluster Sgr~A. 
    Section~\ref{sec:4LSTarray} describes near-future observations of the four-LST array at CTAO-N in the GC using the latest expected instrument response functions together with the methodology developed to forecast for the first time the sensitivity to discriminate among the above-mentioned scenarios with 4 LSTs at CTA-N.
    Section~\ref{sec:4LSTsresults} is devoted to the results and their discussion. 

    \begin{figure}
        \centering
        \includegraphics[keepaspectratio, width=0.8\linewidth, clip]{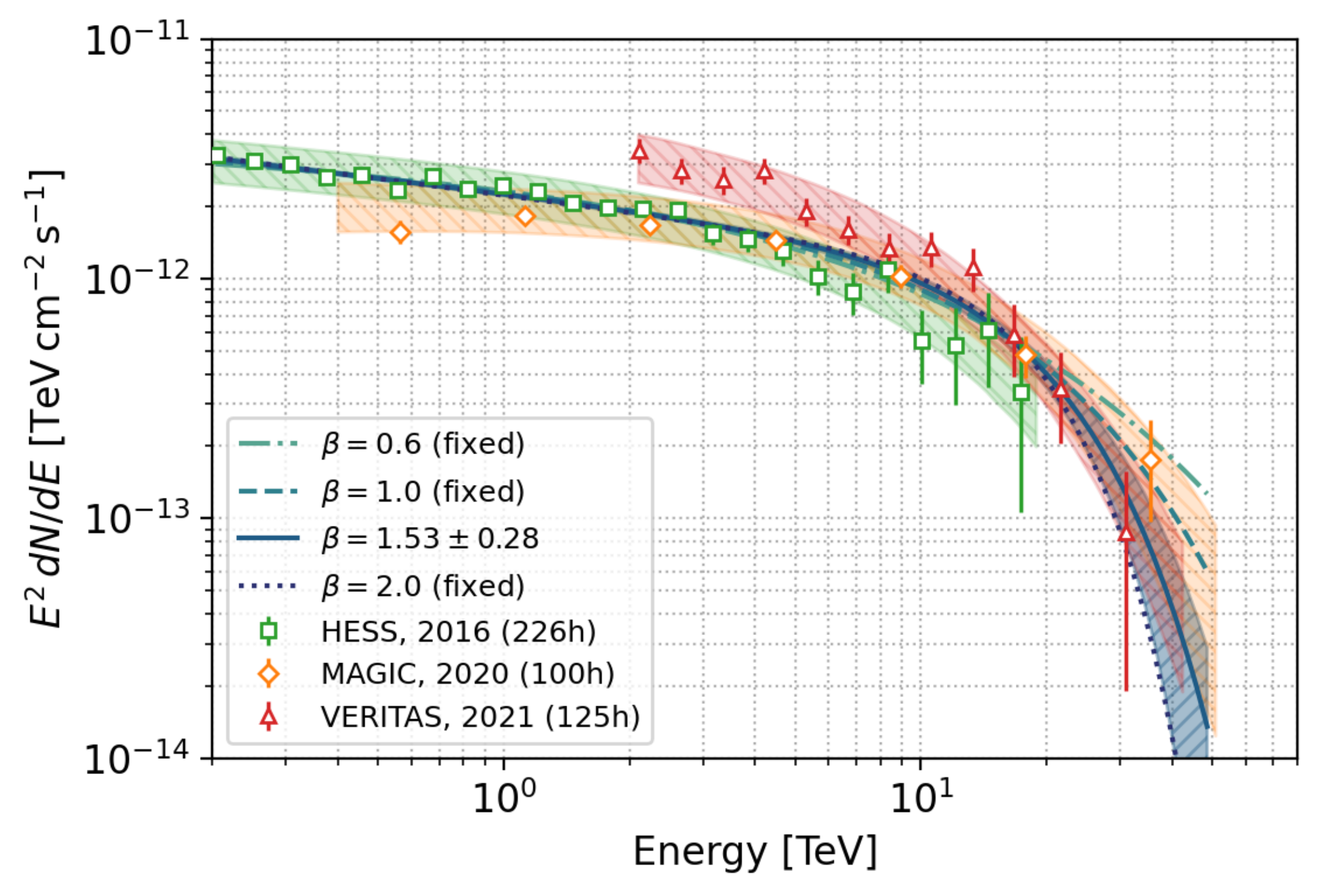}
        \caption{
            Energy-differential flux measurements on the central VHE source from H.E.S.S.~\cite{Abramowski:2016mir} (green), MAGIC~\cite{MAGIC:2020kxa} (orange) and VERITAS~\cite{Adams:2021kwm} (red).
            Data points are taken from the reference publications, and error bars on points represent 68\% statistical uncertainty.
            Error bands correspond to 1$\sigma$ statistical uncertainty convolved with 20\% systematic uncertainty.
            SEPL models are overlaid for the sharpness parameters of $\beta$ = 0.6 (green dotted-dashed line), $\beta$ = 1 (blue dashed line), and $\beta$ = 2 (dotted black line), obtained with the fitted cut-off energy of $E_c = 7.8_{-1.1}^{+1.5} \,\mathrm{TeV}$, $15.7_{-1.5}^{+1.9} \,\mathrm{TeV}$, and $19.2_{-1.5}^{+1.8} \,\mathrm{TeV}$, respectively.
            The blue band is derived from 1$\sigma$ statistical uncertainty of the best-fit model to the three data sets, fitting the sharpness parameter as well. The best-fit parameters are given by  $E_c = 18.6_{-1.6}^{+2.0} \,\mathrm{TeV}$ and $\beta = 1.53 \pm 0.28$.
        }
        \label{fig:J1745spectrum}
    \end{figure}

\section{Spectral models from current VHE gamma-ray data}
\label{sec:models}

    \subsection{Current VHE gamma-ray observations of the GC source}

        The central VHE gamma-ray source has been subject to deep observational programs by 
        current IACTs. Figure~\ref{fig:J1745spectrum} summarizes the latest measurements from H.E.S.S.~\cite{Abramowski:2016mir}, MAGIC~\cite{MAGIC:2020kxa} and VERITAS~\cite{Adams:2021kwm}. 
        A significant softening is measured in the several TeV energy range, providing a fit with an exponential energy cutoff power law (EPL) to the data significantly preferred over a simple power-law one. 
        Following the EPL parametrization obtained, an energy cutoff around 10 TeV is consistently derived from the three experiments within statistical uncertainties. 
        The H.E.S.S. observatory location provides the best visibility of the GC region, which yields a detection threshold of 160 GeV. 
        Observations carried out by MAGIC and VERITAS are performed at higher zenith angles due to their location in the Northern hemisphere, leading to an energy threshold of 400 GeV and 2 TeV, respectively.

        Figure~\ref{fig:J1745spectrum} summarizes the best-fit spectral models measured by the current-generation telescopes H.E.S.S., MAGIC and VERITAS. 
        Each band of the telescope shows a realistic range of the 1$\sigma$ statistical errors convolved with 20\% systematic uncertainty.
        The present photon statistics acquired by current IACTs in the several TeV energy range is limited, which prevent from making any claim on the sharpness of the significant energy cutoff found in the ten TeV energy range. 
        An adequate fit to each dataset  can be obtained for SEPL parametrization with $\beta$ parameter values of 0.6, 1 or 2, which gives energy cutoff values of $E_c = 7.8_{-1.1}^{+1.5} \,\mathrm{TeV}$, $15.7_{-1.5}^{+1.9} \,\mathrm{TeV}$, and $19.2_{-1.5}^{+1.8} \,\mathrm{TeV}$, respectively.
        Performing a $\chi^2$ fit under an assumption of the statistical uncertainty only, the SEPL best fit to the three independent datasets gives $\beta$ = 1.53 $\pm$ 0.28 with the cutoff energy of $E_c = 18.6_{-1.6}^{+2.0} \,\mathrm{TeV}$.
        A fit of the SEPL parametrization to the data including the systematic uncertainties in each of the three current dataset, for which we assume 20\% error of the flux normalization, gives $\beta$ = 1.33 $\pm$ 0.22. 
        Since this study does not aim at discussing the systematic uncertainty budget of the current-generation telescopes, hereinafter we use the statistical-only case, although we lastly evaluate the effect of systematic uncertainties on the LST-array observations to assess realistic performances.
        The present statistic and systematic uncertainty budget shows that SEPL parametrizations provide viable description of the data though they do not enable to further investigate potential underlying emission processes through a conclusive measurement of the sharpness parameter.   
        
       \begin{figure}
            \centering
            \includegraphics[keepaspectratio, width=0.8\linewidth, clip]{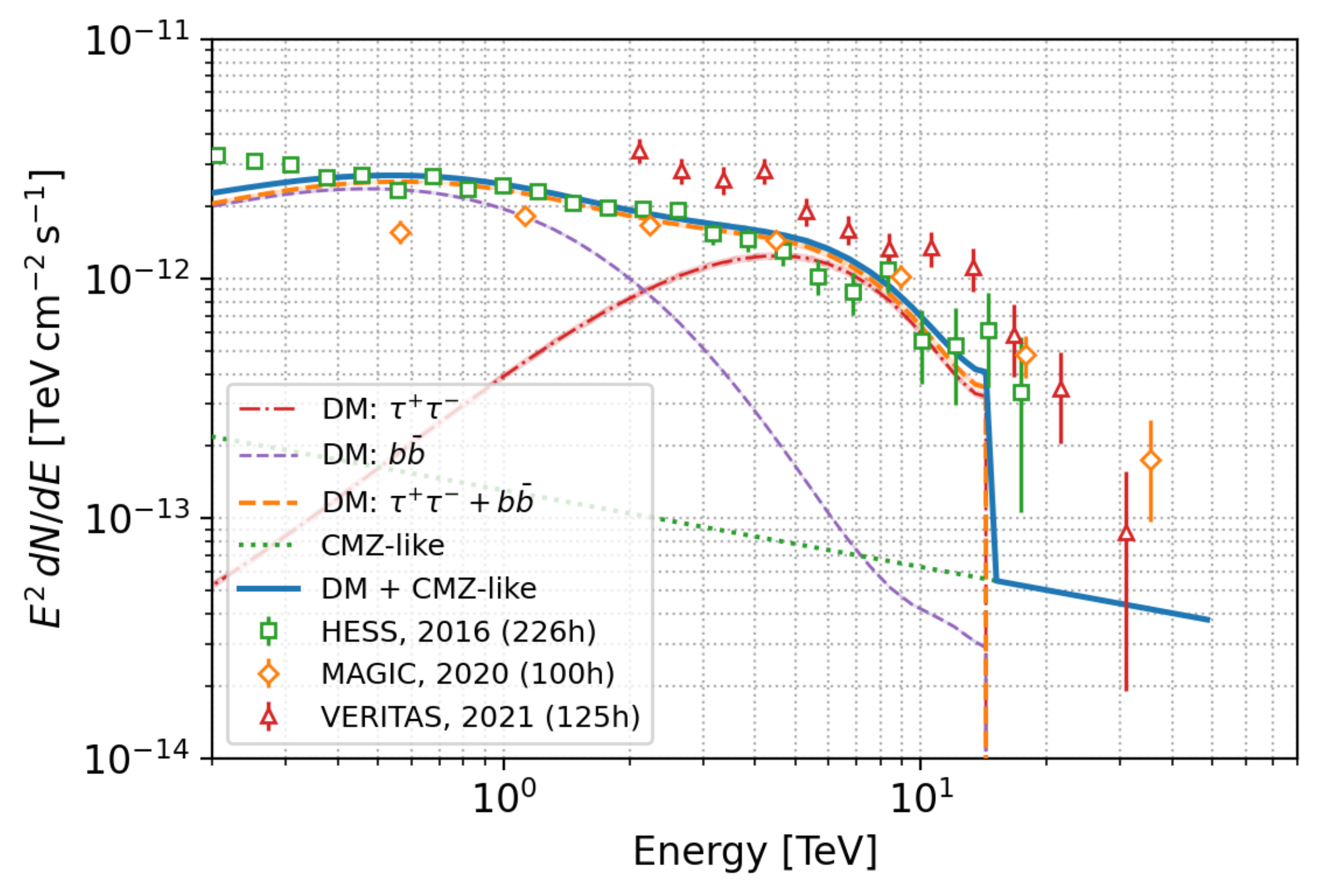}
            \caption{
                Fitted spectral models to the measured energy-differential fluxes measured by H.E.S.S. (green), MAGIC (orange) and VERITAS (red), as for Fig.~\ref{fig:J1745spectrum}.
                The model components correspond to the contribution of $b\bar{b}$ (violet line) and $\tau^+\tau^-$ (red line) channels for the DM annihilation signal for a best-fit DM mass of around 15 TeV, and the contribution of the conventional diffuse emission (CMZ-like, dashed green line).  
                The dashed orange line corresponds to the sum of the 
                $b\bar{b}$ and $\tau^+\tau^-$  contributions. The full model is given by the solid blue line.
            }
            \label{fig:J1745Spectrum_DM}
        \end{figure}

    \subsection{Spike of annihilating dark matter around Sagittarius A*}
    \label{sec:DM}
    
        Motivated by the measurement of an energy cutoff by IACTs, the HESS J1745-290 spectrum has been examined in view multi-TeV annihilating DM~\cite{HESS:2006zwn,Belikov:2013nca}. 
        It has been shown that the interpretation of HESS J1745-290 data if interpreted as DM matter signal, would requires TeV-mass annihilating DM on top of power-law-like emission to fit the low-energy component~\cite{PhysRevD.86.083516,Belikov:2016fwv}.
        The energy-differential flux of gamma rays from the self-annihilation of Majorana particles of mass $m_{\rm DM}$ in a solid angle $\Delta \Omega$ can be expressed as
        \begin{align}
            \frac{d\Phi}{dE}(E_{\gamma}, \Delta\Omega) 
            &= \frac{\langle \sigma v \rangle}{8\pi m^2_{DM}} 
            \sum_i BR_i \frac{dN^i}{dE}(E_{\gamma}) J(\Delta\Omega) 
            \label{eq:flux} \\
            \quad 
            J(\Delta\Omega) 
            &= \int_{\Delta\Omega} \int_{\text{los}} ds\ d\Omega\ \rho^2(s, \Omega),
            \label{eq:J-factor}
        \end{align}
        where $\langle \sigma v \rangle$ stands for the velocity-weighted annihilation cross section and $dN^i/dE$ is the differential yield of gamma rays per annihilation in each channel $i$ with its branching ratio $BR_i$. 
        The term $J(\Delta \Omega)$ in Eq.~(\ref{eq:J-factor}), referred to as $J$-factor, corresponds to the integral of the square of the DM density $\rho$ over the line of sight $s$ and the solid angle $\Delta \Omega$. 
        
        Figure~\ref{fig:J1745Spectrum_DM} presents the overall fitted spectral model to all measured energy-differential fluxes from H.E.S.S., MAGIC and VERITAS. 
        The multi-component model in the inner 0.1$^\circ$ of the GC comprises a mixture of a DM annihilation signal plus a conventional diffuse emission, possibly related to the Central Molecular Zone (CMZ).
        The best-fit description of the present datasets implies a DM annihilation component in mixed $b\bar{b}$ and $\tau^+\tau^-$  final states with branching ratio of BR$_{\tau^+\tau^-}$ = 1 - BR$_{b\bar{b}}$ = $0.34 \pm 0.01$, with a best-fit DM mass of about 15 TeV. 
        Note that a fit to the datasets with a possible 25 TeV DM mass is in agreement with the best-fit model within 1$\sigma$.

        As the expected DM annihilation flux is proportional to the product of the velocity-weighted annihilation cross-section and the J-factor, the DM interpretation of HESS J1745-290 requires a J-factor value of J($<0.1^\circ$) = (2.96 $\pm$ 0.05) $\times$ 10$^{23}$ GeV$^2$cm$^{-5}$ for thermally-produced DM of $\langle \sigma v\rangle$ $\simeq$ 3$\times$10$^{-26}$ cm$^3$s$^{-1}$~\cite{Steigman2012}.
        A high DM concentration around the inner 10 pc of the GC is needed with the above-mentioned J-factor value. While N-body simulations show that the DM density would follow a $1/r$ profile at the center of Milky Way-like galaxies~\cite{Navarro:1996gj}, it has been realized that DM could be adiabatically compressed due the growth of the central black hole~\cite{PhysRevLett.83.1719}, increasing significantly its density.
        The final DM profile and density strongly depend on the initial DM density and profile, the growth history as well as baryon feedback over Gyr timescales~\cite{PhysRevLett.93.061302,PhysRevD.78.083506}. 

        Recent observational progresses on the understanding of the stellar dynamics in the vicinity of Sgr~A* have been made~\cite{2019ApJ...872L..15H,Heissel:2021pcw,2022A&A...657L..12G,2020A&A...634A..71G} and novel determinations of the DM spike density has been derived~\cite{Shapiro:2022prq}.
        The DM spike is embedded in the massive central stellar cluster Sgr A. The gravitational stellar heating of DM in the cluster over Gyr evolution could potentially have softened the DM spike, which therefore could affect the strength of the VHE gamma-ray signal. 
        For the Sgr~A* SMBH of 4.3$\times$10$^6$ M$_\odot$~\cite{GRAVITY:2021xju} and a stellar velocity dispersion in the central stellar cluster extracted for Ref.~\cite{2009ApJ...698..198G}, the heating timescale is a few Gyr such that the stellar heating of the DM spike cannot be excluded. 
        However, as discussed in Ref.~\cite{Merritt:2003qk}, the spike may not have had enough time to relax to the equilibrium profile with a slope of $3/2$. 
        For a recent discussion on the relevant mechanisms of DM formation and dynamical effects at the GC that could have influenced a DM spike around Sgr A*, see, for instance,  Refs.~\cite{Balaji:2023hmy,Zuriaga-Puig:2023imf}.
        Assuming stellar heating in the nuclear cluster, the initial DM spike can be smoothened in the 0.01 pc range only given the few old and bright giant stars below that scale~\cite{2019ApJ...872L..15H,2020A&A...634A..71G}. 
        In that scenario~\cite{Balaji:2023hmy}, a J-factor of a few 10$^{23}$ GeV$^2$cm$^{-5}$ can be obtained for $\langle\sigma v\rangle\sim$ 10$^{-26}$ cm$^3$s$^{-1}$ assuming a conservative value of the local DM density of $\rho_\odot$ = 0.383 GeVcm$^{-3}$~\cite{10.1093/mnras/stw2759}
        \footnote{
            Note, however, that recent GAIA measurements give $\rho_\odot$ = 0.55$\pm$0.17 GeVcm$^{-3}$~\cite{Evans:2018bqy}.
        }.

    \subsection{Proton interaction in the interstellar medium}
    \label{subsec:pp}
    
        Various mechanisms have been proposed to explain the central VHE emission closely connected to the black hole Sgr A*~\cite{Atoyan:2004ix,Aharonian:2004jr,Aharonian:2005ti} or the interaction of protons with the ambient medium in the inner 10 pc~\cite{Liu_2006,Yan-Ping_Wang_2009}. 
        From Ref.~\cite{Chernyakova_2011}, relativistic protons injected with a power-law distribution with an exponential energy cutoff ($\beta$=1), diffuse away from the central source, presumably Sgr A*, and interact in the high interstellar medium density within the few inner pc. 
        Assuming an energy cutoff in the proton spectrum at 100 TeV, the HESS J1745-290 spectrum can be reasonably matched while some degeneracy persists among the model parameters such as the region size, injection history and diffusion coefficient characteristics. 
        In what follows, we will remain agnostic about proton-induced model parameter set and assume a proton distribution following an SEPL spectrum with a spectral index of $\Gamma_p$ = 2.2 and $\beta_p$ = 1. The gamma-ray spectrum produced from $pp$ interaction can be therefore well parametrized by an SEPL spectrum with a spectral index of 2.1 with a shape parameter $\beta \simeq$ 0.6~\cite{2014PhRvD..90l3014K}.

    \subsection{Millisecond pulsars in the central stellar cluster}
    \label{subsec:MSP}

        The central region of the Milky Way harbors a stellar cluster with a mass of 1.5$\times$10$^7$ M$_\odot$ and 0.4 pc core size. 
        Such a cluster may be the result of a merger of several globular clusters due to dynamical friction (see, for instance, Refs.~\cite{2013arXiv1302.2509C,2012ApJ...750..111A,Antonini:2015sza,2020A&ARv..28....4N}). 
        Globular clusters contain more close binary systems per unit mass than the Galactic disk, therefore a higher fraction of millisecond pulsars (MSP)~\cite{2008IAUS..246..291R}. 
        In the recycling scenario, the slowly rotating neutron star accretes mass from a companion, which spins up the neutron star to millisecond period~\cite{1982Natur.300..728A}. 
        On the observational side, MSP has been detected in several Galactic globular clusters such as Terzan 5, 47 Tucan\ae\, and M28. 
        Such globular clusters have been detected as gamma-ray emitters in Fermi-LAT data~\cite{2010A&A...524A..75A}, whose energy spectra show a prominent suppression at a few GeV energies.  
        A massive globular cluster with a mass of 10$^6$ M$_\odot$ could produce close to 100 MSPs~\cite{Ye:2019luh}. Given the nuclear cluster mass, about 1000 MSPs could be harbored in the central nuclear cluster. 
        Interestingly, such a number would provide a cumulative gamma-ray luminosity similar to the luminosity from Fermi-LAT and HESSJ1745-290 measurements assuming a MSP gamma-ray luminosity of $\mathcal{L}_\gamma (> 1\,\rm GeV)\simeq$ 10$^{34}$ erg/s.

        Electrons can be accelerated in MSPs of the central stellar cluster. The escaping pairs injected in the cluster environment diffuse and lose energy via synchrotron and inverse Compton (IC) processes. 
        Several-ten-TeV electrons will lose their energy dominantly via the Inverse Compton scattering off the infrared radiation fields. 
        Following Ref.~\cite{Bednarek:2013oha}, for realistic parameters of MSPs, the interaction of electrons escaping the pulsar magnetosphere could produce VHE gamma rays with a spectrum that matches HESS J1745-290 one, where gamma rays are produced from IC of electrons accelerated to about 100 TeV. The gamma-ray energy cutoff arises from the IC scattering off infrared radiation field of energy of about 0.1~eV~
        \footnote{
            Note that the GeV counterpart can be explained by curvature emission of accelerated electrons along the curving magnetic field lines of the pulsar magnetosphere.
        }.   
        
        In the Klein-Nishina regime, the upscattered Compton spectrum will exhibit the exponential cut-off shape of the parent electron distribution for monochromatic and Planckian target radiation field~\cite{2014ApJ...783..100K}:
        for electron distribution following an exponential energy cut-off power law ($\beta$ = 1), the predicted shape parameter for the spectrum from IC scattering off the Planckian radiation field can be described by a SEPL with a shape parameter $\beta$ =  1.
        Figure~\ref{fig:J1745spectrum} shows the VHE gamma-ray emission from IC scattering off soft infrared and optical background radiation by relativistic leptons escaping the population of MSPs located in the central stellar cluster that well fit to the HESS J1745-290 data. 
        Escaping electrons are injected in the central cluster following EPCL with a spectral index 2 and an energy cutoff $E_{\rm cut}^{\rm e}$ of 50 TeV as expected for typical MSP parameters~\cite{Bednarek:2013oha}.

\section{The Galactic Center with four Large-Sized Telescopes at CTAO-N}
\label{sec:4LSTarray}

    \subsection{Galactic Center observations}
    
        The Cherenkov Telescope Array Observatory is the next-generation ground-based observatory for gamma-ray astronomy at very high energies~\cite{cta_observatory,CTAConsortium:2017dvg}. 
        CTAO will cover a wide energy range, from 20 GeV to 300 TeV, by three types of telescopes with different diameters. Each telescope is called the Large-Sized Telescope (LST), the Medium-Sized Telescope (MST) and the Small-Sized Telescope (SST) when ordered by dish size. 
        CTAO will consist of two observatories for full sky coverage. The southern one (CTAO-South) is planned to be at Paranal in Chile (-24$^{\circ}$S 37$'$, -70$^{\circ}$W 24$'$), while the northern observatory (CTAO-North) is located at the Roque de los Muchachos Observatory in the Canary Island of La Palma (28$^{\circ}$N 45$'$, 18$^{\circ}$W 53$'$).

        The performances of IACTs heavily depend on the zenith angles (ZA) at which they would perform observations since this would change the distance to an air shower maximum as well as atmospheric depth.
        In this paper, we focus on the case of forecast LZA observations of the GC with the four-LST subarray in CTAO-North given the visibility window of the GC ant thta location. 
        The Instrumental Response Functions (IRF), including energy-dependent effective area, angular and energy resolutions, are computed from Monte Carlo simulations and taken from the \textit{publicly available} \texttt{North-LSTSubArray-60deg} from \texttt{prod5-v0.1} library \cite{ctao_2021_5499840}, using 42 logarithmically-spaced energy bins from 20 GeV to 200 TeV.
        Those energy and angular resolutions in IRFs are defined as 80\% containment and are estimated from 100 GeV, each of them is 10\% and 12\%, respectively, which indicates, in particular, this LZA observation is the optimized setup for energies at several ten TeV.

        With the location of CTA-North observatory, the central region of the Milky Way can be observed from April to August each year with zenith angles higher than 58$^\circ$. 
        The right-ascension band of GC visibility contains a broad range of astrophysical objects. 
        Accounting for the prioritization of different Right Ascension objects in such a band, at most 100 hours of observations near the GC can be realistically obtained per year. 
        We assume a mean zenith angle of 60$^\circ$ for the observations as a realistic value considering the various constraints in this visibility window for the CTAO-N site. 
        We consider data collected in stereo mode using the four LSTs, and use the appropriate instrument response functions at 60$^\circ$ extracted from Ref.~\cite{ctao_2021_5499840}.
        It leads to accessible energies up to 100 TeV, setting 400 GeV as a conservative threshold, though we also consider bringing it down to 250 GeV.  
        \begin{figure*}[!ht]
            \begin{minipage}[b]{0.49\textwidth}
                \centering        \includegraphics[width=0.95\columnwidth]{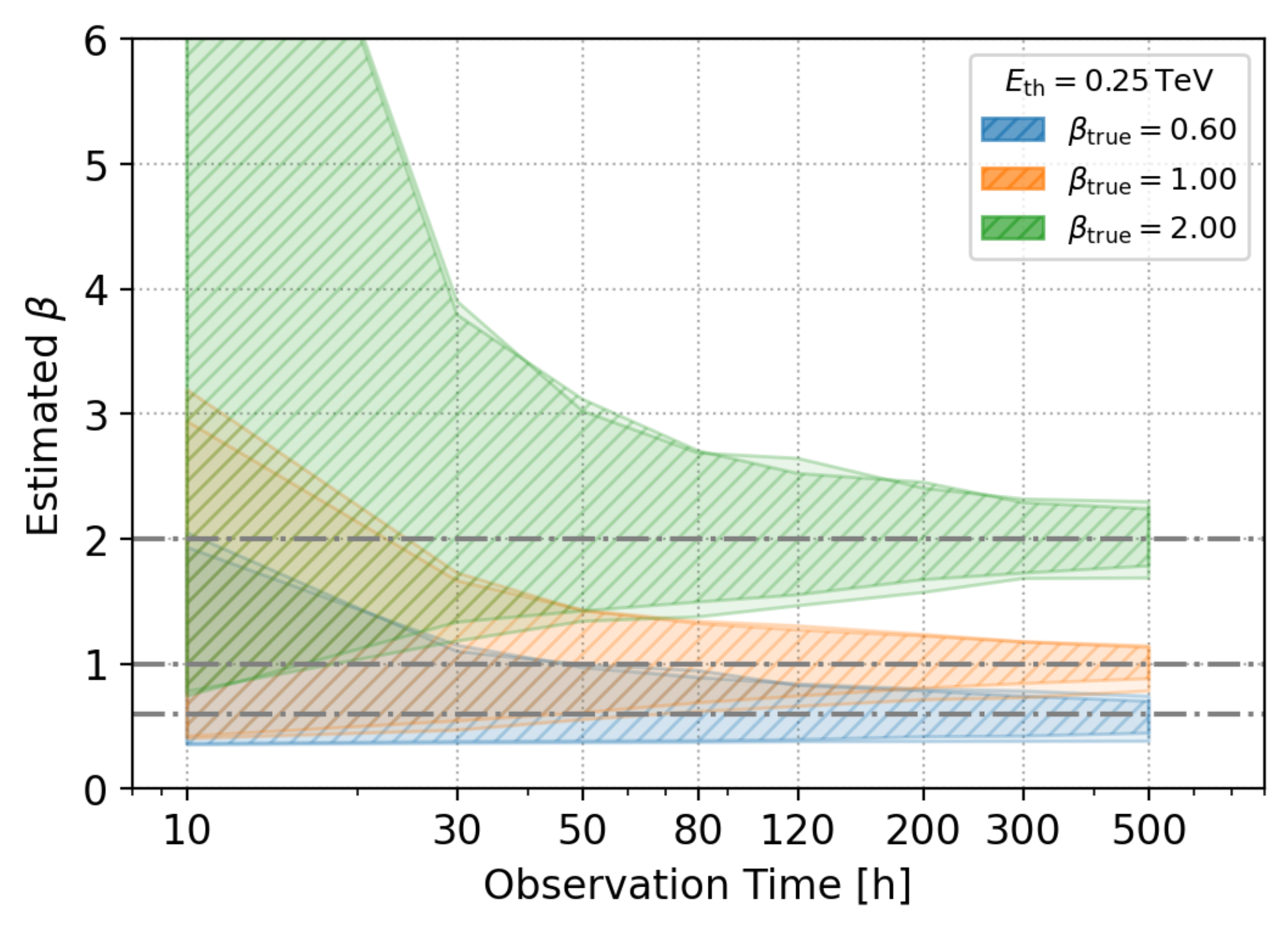}
            \end{minipage}
            \begin{minipage}[b]{0.49\textwidth}
                \centering
                \includegraphics[width=0.95\columnwidth]{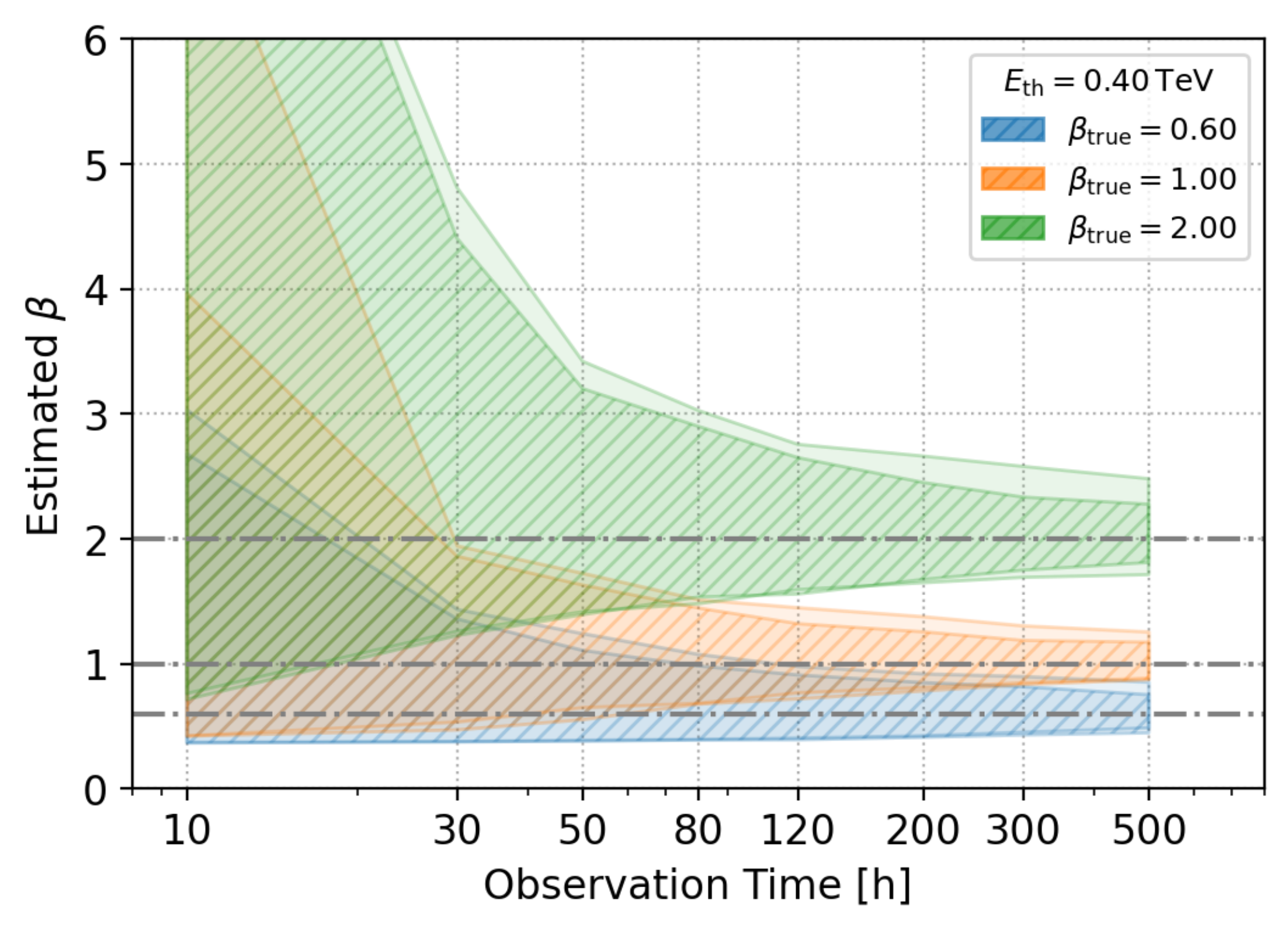}
            \end{minipage}
            \caption{
                Time evolution of the 68\% containment of the derived $\beta$ parameter, computed by fitting the models to the data sets that are simulated by the exponential templates.
                The fit was performed assuming two different energy thresholds: $E_{\rm th} = 0.25$ TeV (left panel) and $E_{\rm th} = 0.40$ TeV (right panel). 
                The hatched and not-hatched bands show the statistical-uncertainty-only and systematic-uncertainty-included cases, respectively. Horizontal dash-dotted lines show the assumed underlying $\beta$ parameters.
            }
            \label{fig:estimated_beta}
        \end{figure*}

    \subsection{Mock Data Production and Analysis}

        We simulated observations of the central source HESS J1745-290 with the 4-LST subarray through \texttt{gammapy v1.1}, making use of IRFs publicly available from \texttt{prod5-v0.1} library~\cite{ctao_2021_5499840} based on 50 hours of observations carried out with the south pointing at a zenith angle of 60$^\circ$. 
        As with the current operation of LST-1~\cite{CTALSTProject:2023wzu}, this study assumed wobble-mode observations with an offset angle of 0.7$^\circ$ around Sgr A*, with the pointing positions aligned to the Galactic coordinates. 
        A count cube $N(x, E)$ as a function of the reconstructed arrival direction $x = (l, b)$ and the reconstructed energy $E$ is associated with models $\Phi(\bar{x}, \bar{E})$ through the instrument response function $R(\bar{x}, \bar{E} \mid x, E)$:
        \begin{align}
            N(x, E) 
            =  
            \int {\rm d} \bar{E} \, 
            \int {\rm d} \bar{x} \
            R(x, E \mid \bar{x}, \bar{E}) 
            \times 
            \Phi(\bar{x}, \bar{E}),
            \label{eq:folding}
        \end{align}
        where $\bar{x}$ and $\bar{E}$ are the true arrival direction and the true energy, respectively. 
        
        The response function $R$ is defined as the product of three independent functions:
        \begin{align}
            R 
            = \varepsilon \left( \bar{x}, \bar{E} \right) 
            \times {\rm PSF} \left( x \mid \bar{x}, \bar{E} \right) 
            \times E_{\rm disp} \left( E \mid \bar{x}, \bar{E} \right),
            \label{eq:IRF}
        \end{align}
        where $\varepsilon = t_{\rm obs} \times A_{\rm eff} \left( \bar{x}, \bar{E} \right)$ is the exposure defined as the product of the observation time and the effective area. ${\rm PSF}$ stands for the point spread function, and $E_{\rm disp}$ is the energy dispersion. Without an analytic formula, those original dispersion shapes are used for convolution for IRF calculation directly, following the standard 3D analysis in \texttt{gammapy}.
        Here, no additional gamma-ray source is presumed within the FoV. Namely, models comprise two components: a point source at the position of Sgr A* and residual background resulting mainly from misidentified cosmic rays. 
        The residual background is effectively assumed to be subtracted either from measurements in the empty regions within the field of view, such as using \textit{RingBackground} or \textit{Reflected-Region-Background} methods~\cite{Berge:2006ae}, or via modeling using dedicated Monte Carlo simulations for the given observational and instrumental conditions~\cite{Holler:2020duc}.

        Using the templates defined in Sec.~\ref{sec:models}, count cubes are realized through Eq.~\eqref{eq:folding} with the statistical uncertainty, \textit{i.e.}, following the Poisson statistics. This study adopted a typical cube geometry: $0.02^{\circ} \times 0.02^{\circ}$ spatial bins and 5 energy bins per decade (0.2 dex).
        The gamma-ray signals are formulated by performing a fit of spectral and spatial models to the simulated count cubes through the forward-folding approach. The Cash statistics~\cite{Cash:1979ApJ} is adopted to evaluate the fit to the data, thereby not reflecting uncertainties related to the response functions at the likelihood level. 
        In this study, any parameter is unfrozen in neither the spatial components nor the background model, while the spectral component of Sgr A*, given in Eq.~\eqref{eq:general_ECPL}, are optimized through the fit. The fit is performed with all pixels within the wobble distance from Sgr~A*, as the pixels outside are assumed to be used for a background estimation (see, e.g., Ref.~\cite{CTALSTProject:2023wzu}), albeit in this study the background is modeled directly from the response function.

        The LZA observation generally entails higher systematic uncertainties, on account of the telescope response more sensitive to the zenith angle and the longer travel in the more rarefied atmosphere. 
        For instance, the past research with the MAGIC telescopes reported a 15\% (10 \%) systematic uncertainty on the energy scale and $\sim$20 \% (15 \%) for the flux normalization \cite{Ahnen:2016crz, MAGIC:2020kxa, MAGIC:2022acl,2015PhDT176F}, considering the LZA (typical low ZA) observation setup. Such uncertainties are comparable to H.E.S.S. where 20\% is assumed on the flux normalisation and 10\% on the energy scale.
        While the statistical uncertainty is mainly evaluated in the previous findings~\cite{Belikov:2016fwv}, this study additionally estimates systematic effects resulting from misestimated responses, and assess how these uncertainties possibly impede the determination of the energy spectrum. 
        To mock the effects, in this study, four types of response uncertainty are artificially introduced into the response functions when simulating the count cubes, whereas the official responses are used when the fit undergoes:
        \begin{itemize}
            \item 
            $\pm$ 15\% shift of the collection area: 
            $
                \tilde{A}_{\rm eff} \left( \bar{x}, \bar{E} \right) 
                = A_{\rm eff} \left( \bar{x}, \bar{E}/(1+s_1) \right) 
            $,
            where the additional scale parameter $s_1$ is randomly generated from the Normal distribution whose standard deviation is 0.15.
            \item 
            $\pm$ 15\% shift of the energy scale: 
            $
                \tilde{E}_{\rm disp} \left( E \mid \bar{x}, \bar{E} \right)
                = 
                E_{\rm disp} \left( E/(1+s_2) \mid \bar{x}, \bar{E} \right)
            $,
            where the additional scale parameter $s_2$ is randomly generated from the Normal distribution with standard  deviation of 0.15;
            \item 
            $\pm$ 15\% widening of the energy resolution: 
            $
                \tilde{E}_{\rm disp} \left( E \mid \bar{x}, \bar{E} \right)
                = 
                E_{\rm disp} \left( {\mathcal N} (E) \mid \bar{x}, \bar{E} \right)
            $, 
            where the smearing function ${\mathcal N}$ is the Normal distribution with standard deviation of 0.15;
            \item 
            $\pm$ 1\% amplitude scale of the background model;
            $
                \tilde{\Phi}_{\rm bkg} \left( E \mid \bar{x}, \bar{E} \right)
                = 
                (1 + s_4) \cdot \Phi_{\rm bkg} \left( E \mid \bar{x}, \bar{E} \right)
            $,
            where the additional scale parameter $s_4$ is randomly generated from the Normal distribution whose standard deviation is 0.01.
        \end{itemize}  
        
\begin{figure}[!ht]
    \centering

    \begin{minipage}[t]{0.48\columnwidth}
        \centering
        \includegraphics[width=\linewidth]{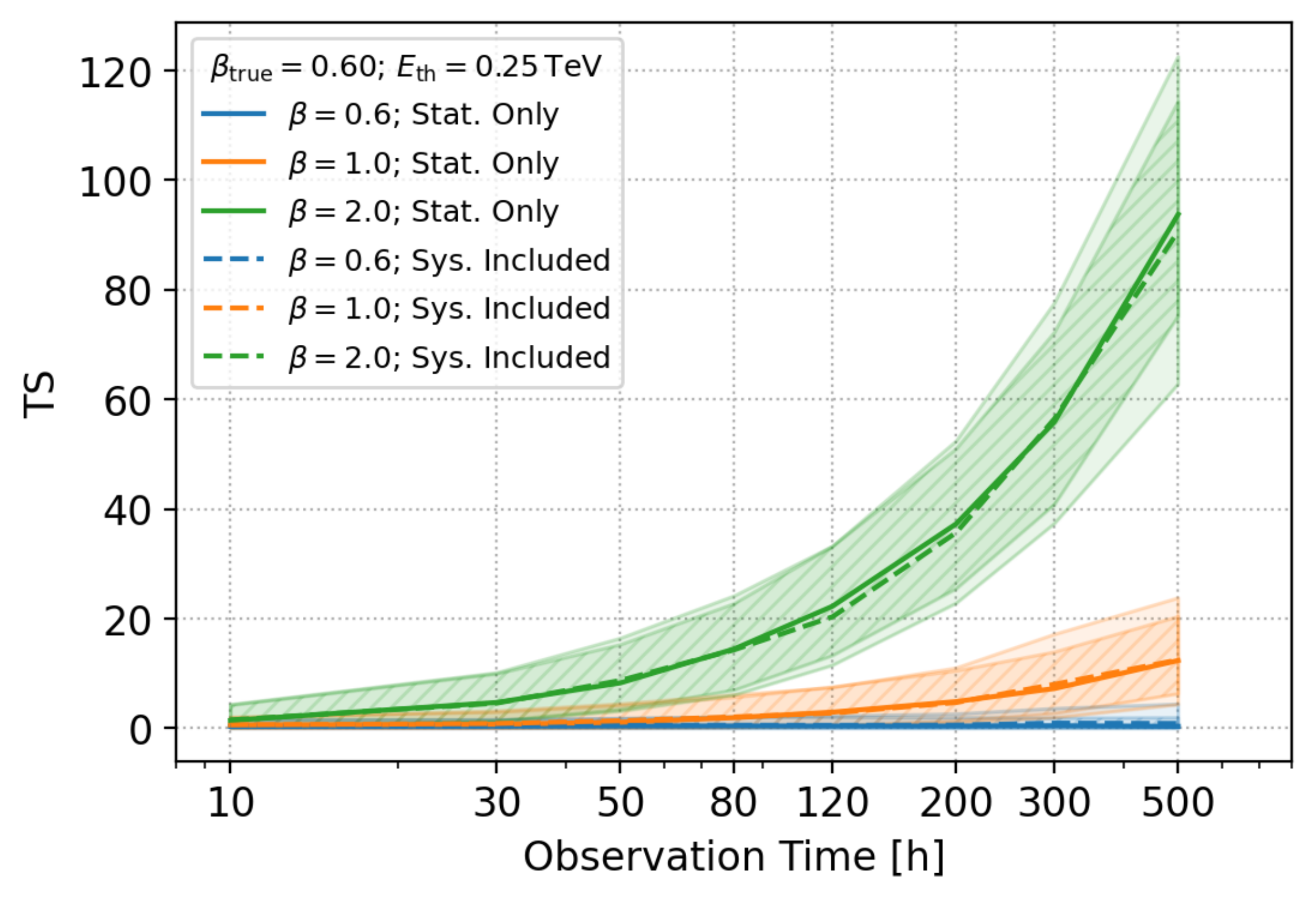}
    \end{minipage}
    \hfill
    \begin{minipage}[t]{0.48\columnwidth}
        \centering
        \includegraphics[width=\linewidth]{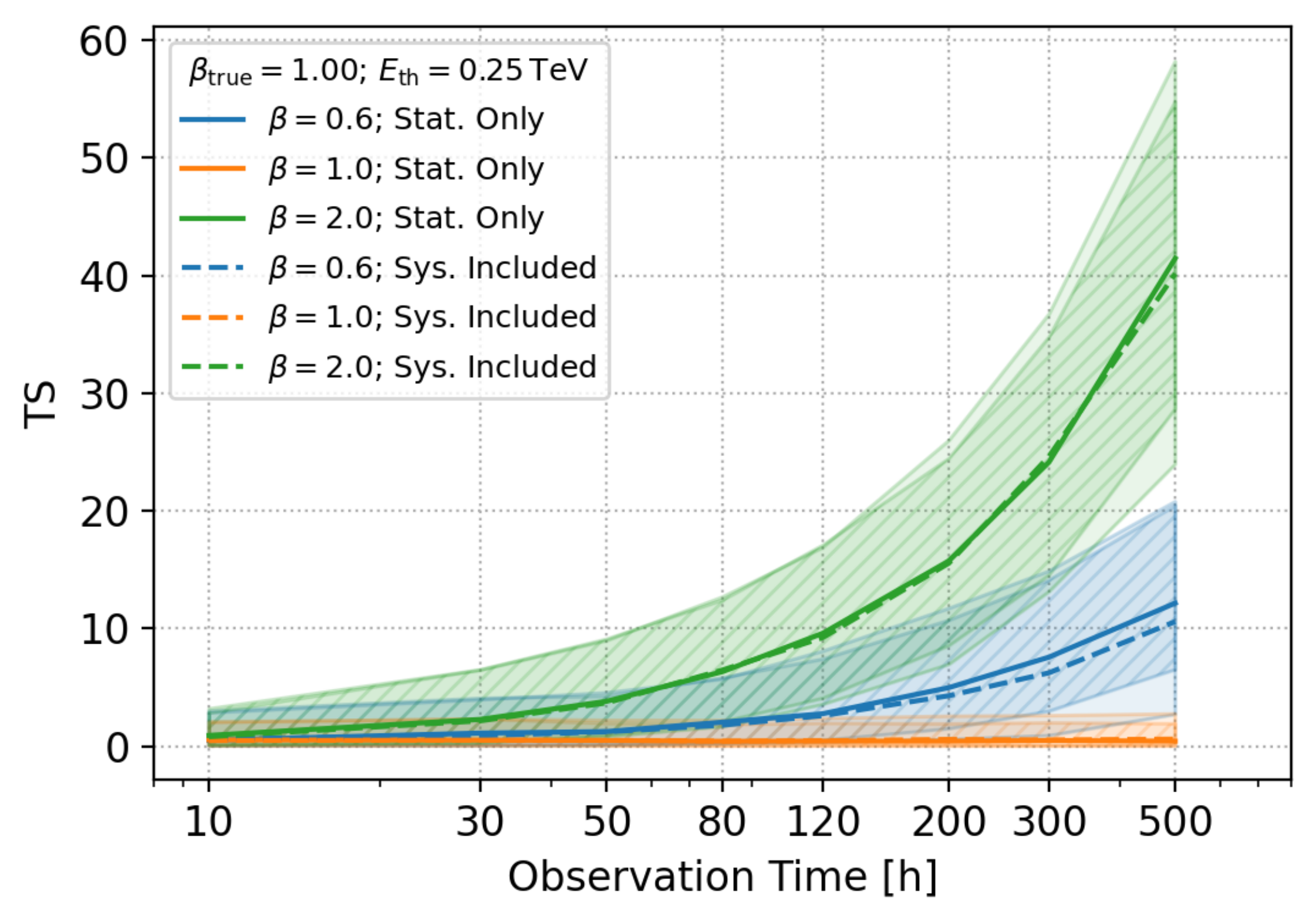}
    \end{minipage}
    \vspace{0.5em}
    \begin{minipage}[t]{0.48\columnwidth}
        \centering
        \includegraphics[width=\linewidth]{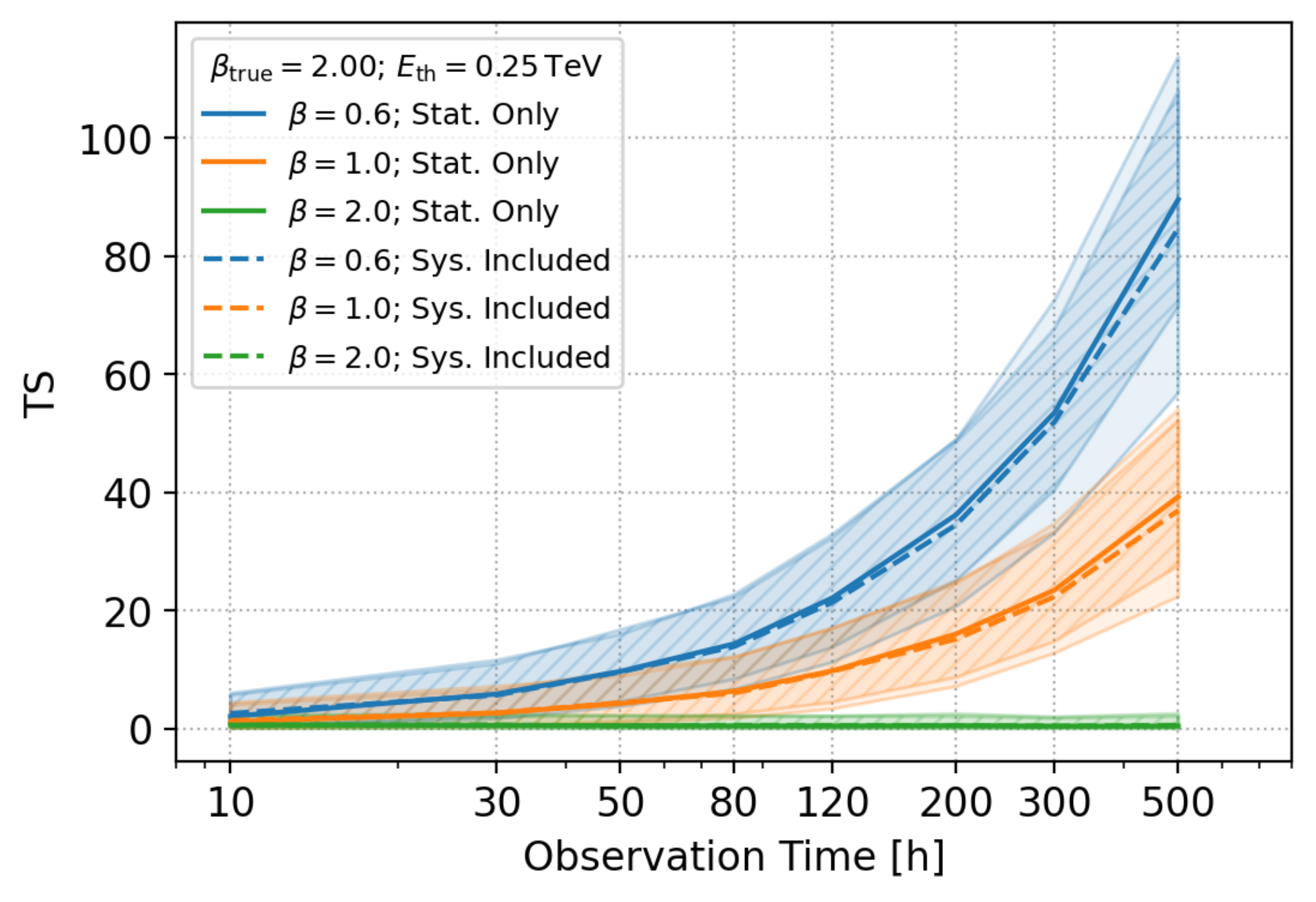}
    \end{minipage}
    \caption{
        Test statistics (TS) values as a function of observation time for true $\beta$ parameter values of 0.60 (top left), 1.00 (top right), and 2.00 (bottom), respectively.
        In each case, the fit to the data is performed assuming different $\beta$ parameter values for an energy threshold of 0.25 TeV. 
        The hatched and not-hatched bands show the 68\% containment for the statistical-uncertainty-only and systematic-uncertainty-included cases, respectively. 
        Lines within the bands show the TS median value.
    }
    \label{fig:ts_beta}
\end{figure}

\section{Results and discussion}
\label{sec:4LSTsresults}
    
    Figure~\ref{fig:estimated_beta} shows distributions of the estimated $\beta$ parameter as a function of the observation time, for analysis energy thresholds of 0.25 TeV (left panel) and 0.40 TeV (right panel). 
    This result indicates that the discrimination power between the scenarios especially between $\beta = 1.0$ and $\beta = 2.0$ significantly increases with accumulated data.
    Furthermore, the addition of the systematic uncertainties does not strongly affect the parameter estimation, widening the 68\% containment band by ~20\% at $t_{\mathrm{obs}} = 120\,\mathrm{hr}$ and ~40\% at $t_{\mathrm{obs}} = 500\,\mathrm{hr}$. 
    Notably, systematic uncertainties have a greater impact at lower energy thresholds, but this can be mitigated by using a more conservative threshold, as done in this study.

    In order to statistically test the discrimination power among the three viable models with accumulated photon statistics through increased observation time, the likelihood ratio is computed:
    \begin{align}
        \mathrm{TS} 
        = -2 \left( 
            \ln {\mathcal L} \left( \beta = \beta_{\mathrm{fixed}} \right)
            - \ln {\mathcal L} \left( \beta = \beta_{\mathrm{fitted}} \right)
        \right),
    \end{align}

    TS  greater than 9 and 25 respectively corresponds with the significance at the $3\sigma$ and $5\sigma$ confidence levels, according to Wilks' theorem \cite{Wilks:AMS:9:1938}. 
    We evaluate the TS distributions from about 800 realizations of data and analysis at each setup, \textit{i.e.}, the observation time and the spectral sharpness parameter $\beta$.

    Figure~\ref{fig:ts_beta} shows TS values of a fixed $\beta$ parameter of 0.6, 1.0 and 2.0, respectively, versus observation time, for a given true $\beta$ value, $\beta_{\rm true}$, assuming an energy threshold of E$_{\rm th}$ = 0.25 TeV.
    For $\beta_{\rm true}$ = 0.6 (top panel), the $\beta$ = 2.0 scenario can be rejected at 4.7 and 9.7$\sigma$ for 120 and 500 hours, respectively, while the $\beta$ = 1.0 case is only rejected 1.7$\sigma$ even for 500 hours.
    For $\beta_{\rm true}$ = 1.0 (middle panel), the case of $\beta = 2.0$ can be also rejected at 3.1$\sigma$ for 120 hours, whereas the $\beta$ = 0.6 scenario requires 500 hours to reach the $\sim$3$\sigma$ level.
    For $\beta_{\rm true}$ = 2.0 (bottom panel), the case of $\beta = 0.6$ is rejected at 4.7$\sigma$ for 120 hours, and the $\beta = 1.0$ one is also significantly ruled out with the 500 hour observations.
    In general, the $\beta$ values that differ from more that one unity compared to the value of $\beta_{\rm true}$ can significantly rejected with 500 hours of observation time.

\section{Summary}

    The power-law spectral model with either exponential or super-exponential cutoff can adequately describe the present data of the VHE gamma-ray emission from the inner 10 parsecs of the GC region, measured with the three current-generation Cherenkov telescopes, H.E.S.S., MAGIC, and VERITAS. 
    Such paramatrizations of the present spectral measurements can be provided by physically-motivated emission models, such as spike of DM annihilation around Sgr~A*, proton interaction in the interstellar medium, or millisecond pulsar emission in the central stellar cluster. 
    Further discrimination between these models can be obtained by an accurate determination of the spectral cutoff sharpness $\beta$.
    Nonetheless, the presently-available statistics with the systematic uncertainties do not enable to significantly discriminate among the above-mentioned scenarios.

    Given the imminent deployment of the CTAO-N four-LST sub-array by 2026, we provide a timely and realistic simulation of upcoming observations of the GC region, using the latest knowledge of the IRFs. 
    This study derives for the first time its sensitivity forecast to resolve the sharpness of the spectral energy cutoff by modeling the expected dataset with exponential-cutoff power-law models.
    With approximately 800 realizations for each simulation setup, we evaluate the uncertainty in the sharpness parameter determination, and quantify how much four-LST observations at CTAO-N enable to discriminate among the scenarios discussed in Sec.~\ref{sec:DM}.
    We consider not only up to about 100 hours of observations assumed to be feasibly conducted within the first year, but also extend the simulations up to 500 hours to evaluate the performances of a potential multi-year campaign.

    We demonstrate that observations with the forthcoming four-LST array at the CTAO-N site, enhanced by the large-zenith-angle observation technique, provide valuable insights in discriminating the value of the sharpness parameter of the energy cutoff.
    Approximately 500 hours of observation will provide a conclusive discrimination, at more than 5$\sigma$ level, between the DM annihilation spike model and the two other alternative scenarios.
    In addition, with 120 hours of observations, feasibly taken within the first year of operation of the four-LST array, spectral measurements can discriminate between a DM annihilation spike model ($\beta$ = 2) and a MSP one ($\beta = 1$) at about 3$\sigma$ level.
    While the CTAO-S array is expected to provide an unprecedented view of the GC region, the four-LST sub-array at CTAO-N, expected to begin operations in 2026, can offer a timely opportunity to further probe the VHE gamma-ray emission in the inner 10 parsecs of the GC.

    Additionally, while our study primarily considers four LSTs, the future deployment of MSTs at the CTA-N array will further enhance its observational capabilities on the GC.
    The primary goal of this study is to assess, for the first time, how a 4-LST sub-aarra at CTAO-N can contribute in the next several years to distinguishing different gamma-ray emission scenarios. 
    Our sensitivity study is crucial to robustly quantify expectations of near-future measurements. 
    The CTAO Southern site array will unambiguously enable superior capabilities for studying HESS J1745-290 compared to CTA-N. 
    With the deployment of 2 MSTs and 5 SSTs within a few years, new insights on the orgin of HESS J1745-290 are expected given the ideal location of CTA-S to observe the GC under the best conditions.
    Four LSTs at CTAO-N will be operational in 2026 and we highlight here how an observational program at CTA-N with about 100 hours taken per year can provide scientific opportunities in the interim period.

\acknowledgments
This work was conducted in the context of the CTAO Consortium. 
We appreciate useful comments in the CTAO internal review process. 
It has made use of the CTAO instrument response functions provided by the CTAO Consortium and Observatory. 
This work of S.A. was supported by JST SPRING, Grant Number JPMJSP2108, and by Grant-in-Aid for JSPS Fellows, Grant Number 24KJ0544.
This work of T.I. was supported by JSPS KAKENHI Grant Number 20KK0067.

\newpage
\bibliographystyle{JHEP}
\bibliography{bib.bib}
\end{document}